\documentclass[oldversion]{aa}
\usepackage{graphicx}
\usepackage{longtable}
\begin{document}

\title{Radio properties of H$_{2}$O maser host galaxies}

\author{J.S.~Zhang\inst{1} \and  C.~Henkel \inst{2,3} \and  Q. Guo \inst{4} \and J.~Wang \inst{1}}
\offprints{J.S. Zhang, jszhang@gzhu.edu.cn}

\institute{ Center For Astrophysics, GuangZhou University, GuangZhou, 510006, China
       \and
           Max-Planck-Institut f{\"u}r Radioastronomie, Auf dem H{\"u}gel 69, D-53121 Bonn, Germany
       \and
           Astronomy Department, King Abdulaziz University, P.O. Box 80203, Jeddah 21589, Saudi Arabia
       \and
           Hunan Institute of Humanities, Science and Technology, Loudi, 417000, China
       }

\date{received ; accepted}

\abstract{The 6\,cm and 20\,cm radio continuum properties of all 85 galaxies
with reported 22\,GHz H$_2$O maser emission and luminosity distance $D$$>$0.5\,Mpc
are studied. For the total of 55 targets for which both 6\,cm and 20\,cm measurements
exist and for the subsample of 42 sources with masers related to active galactic
nuclei (AGN), a spectral index could be determined from an assumed power-law dependence.
The mean value of the resulting spectral index is in both cases $0.66\pm0.07$
($S$$\propto$$\nu^{-\alpha}$; $S$: flux density, $\nu$: frequency). Comparing
radio properties of the maser galaxies with a sample of Seyferts without detected
H$_2$O maser, we find that (1) the spectral indices agree within the error limits, and
(2) maser host galaxies have higher nuclear radio continuum luminosities, exceeding
those of the comparison sample by factors of order 5. Only considering the subsample
of galaxies with masers associated with AGN, there seems
to be a trend toward rising maser luminosity with nuclear radio luminosity
(both at 6\,cm and 20\,cm). However, when accounting for the Malmquist effect, the correlation
weakens to a level, which is barely significant. Overall, the study indicates that
nuclear radio luminosity is a suitable indicator to guide future AGN maser searches
and to enhance detection rates, which are otherwise quite low ($<$10\%).}

\keywords{Masers -- galaxies: active -- galaxies: nuclei -- radio lines: galaxies --
radio continuum: galaxies}

\titlerunning{Radio properties of H$_{2}$O maser galaxies}
\authorrunning{J.S. Zhang et al.}

\maketitle

\section{Introduction}

Since the first detection of an extragalactic H$_2$O maser towards M\,33 (Churchwell
et al. 1977), 22\,GHz ($\lambda$=1.3\,cm) H$_2$O maser surveys have covered $\sim$3000
galaxies, with reported detections from 85 objects (e.g., Zhang et al. 2006; Braatz \&
Gugliucci 2008; Greenhill et al. 2008; Darling et al. 2008). With respect to their
physical origin, these maser sources can be subdivided into two classes: (1) those
associated with star-forming regions (hereafter SF-masers), which are similar to the
stronger sources in the Galaxy, and (2) those associated with active galactic nuclei
(hereafter AGN-masers). The majority of the extragalactic H$_2$O masers are believed
to belong to the latter class, mainly because among the small number of interferometrically
studied sources, all of the masers with isotropic luminosities $>$10\,L$_{\odot}$ are
found to be associated with AGN. With their exceptional luminosities, far surpassing
those of any galactic maser, these molecular lighthouses are termed ``megamasers''.
Among AGN maser sources, a large portion ($\sim$40\%) has been identified as disk maser
candidates (Kondratko et al. 2006; Zhang et al. 2010). Their maser spots are located
within the central few pc, forming parts of a molecular accretion disk around the nuclear
engine. In addition to ``systemic'' features, ``high velocity'' components are also found,
representing the approaching and receding edges of the disks, which are viewed approximately
edge-on. Systematic studies of such disk masers have become a promising tool for addressing a
wide variety of astrophysical problems, using both high resolution VLBI (Very Long Baseline
Interferometry) imaging and spectral line monitoring of maser spots (Braatz et al. 2009).

Observations show that H$_{2}$O megamasers are mostly associated with the nuclear regions
of Seyfert 2 and LINER (low ionizing nuclear emission regions) galaxies, which are commonly
heavily obscured (gas column density $N_{\rm H}$$>$10$^{23}$cm$^{-2}$) (Braatz et al. 1997;
Zhang et al. 2006; Greenhill et al. 2008). AGN are considered to be the ultimate energy
source of H$_{2}$O megamasers (e.g., Lo 2005), so we may expect some kind of
correlation between maser power and the nuclear radio luminosity (for its definition, see
Sect.\,2.1). On the one hand, the nuclear radio continuum luminosity has been
established as a useful isotropic luminosity indicator of AGN power (Giuricin et al. 1990,
Diamond-Stanic et al. 2009). On the other, H$_{2}$O line emission may be produced by
amplifying the nuclear radio emission, which can provide ``seed" photons for masers
located on the front side of the nucleus or an associated nuclear jet (Braatz et al. 1997;
Henkel et al. 1998).

In this paper, we compile radio measurements from the literature for the entire sample of
reported H$_{2}$O maser host galaxies with luminosity distances $D$ $>$
0.5\,Mpc in order to investigate their radio properties and probe possible correlations between
H$_{2}$O megamasers and the nuclear radio emission. The maser sample, as well as an unbiased
comparison sample with no detected 22\,GHz H$_2$O emission but good upper limits, is
presented in section 2. In section 3, we analyze their radio properties. Their properties are compared,
emphasizing 6\,cm and 20\,cm nuclear radio flux distributions and spectral
indices. Furthermore, correlations between maser
emission and nuclear radio power are explored. These results are summarized in
section 4.

\section{Samples and data}

\subsection{The H$_2$O maser sample}

All 85 galaxies with luminosity distances $D$ $>$ 0.5\,Mpc and reported H$_{2}$O maser
emission are presented in Table\,2. They include 66 AGN-masers, ten SF-masers (isotropic luminosity
$L_{\rm H_{2}O}<10\,L_{\odot}$, known as ``kilomasers"), and additional low-luminosity
sources, which are still awaiting interferometric observations to identify their nature (for
details of their classification and activity types, see Zhang et al. 2010). For radio
continuum measurements of this maser galaxy sample, we have used the NASA Extragalactic
Database (NED; http://nedwww.ipac.caltech.edu), the NVSS (NRAO VLA Sky Survey), and FIRST (Faint
Images of the Radio Sky at Twenty-centimeters) catalogs. Since H$_{2}$O maser galaxies are
mostly Seyfert 2s or LINERs, their radio continuum tends to be weak and is dominated by the
circumnuclear region (Braatz et al. 1997). Most sources have been measured at 6\,cm and 20\,cm,
while just a few sources have been observed at other wavelengths. Therefore we have collected
the 6\,cm and 20\,cm data and use them as nuclear radio output.

The nuclear radio properties of maser host galaxies were already studied by Braatz et al. (1997).
There were only 16 maser sources known at that time and just nine sources with available radio
continuum data, inhibiting any detailed statistical analysis. Therefore, the current maser
sample provides a drastic improvement. In Braatz et al. (1997), 6\,cm VLA data measured in the
A and B configurations were taken to ensure that only the small-scale flux was considered. For
our current larger sample, there are no VLA observations for some sources and many sources were
observed by different telescopes with different beam size. When multiple data were available,
measurements were selected according to the following criteria
 (1) Observations with smaller beam sizes are preferred, in order to isolate the nuclear from the large scale emission of the host galaxy
(Diamond-Stanic et al. 2009). This might cause missing flux when selecting interferometric data.
However, the best VLA resolution at 6\,cm is $\sim$0.4\,arcsec, which corresponds to a linear
size of $\sim$20\,pc for a source at distance 10\,Mpc. This size and the corresponding highest
VLA resolution at 20\,cm, $\sim$2\,arcsec, is much larger than that of the maser spots observed
on sub-pc scales (e.g., Miyoshi et al. 1995, Reid et al. 2009) and that of potentially associated
nuclear continuum sources.

(2) Both 6\,cm and 20\,cm data should have similar beam size. The most
stringent requirement would be that data for a given source were obtained from high resolution
VLA observations. Six centimeter data from the B configuration and 20\,cm data from the A configuration
would be optimal.

Overall, we got 6\,cm data for 57 sources and 20\,cm data for 79 sources. There are 55 sources
with 6\,cm {\it and} 20\,cm data, including 41 sources with both flux densities from the same telescope.
For those 41 sources, 27 have been measured with the VLA at both frequencies. Among these, 11
belong to the ideal case, with A-array 20\,cm and B-array 6\,cm data. For those sources with
6\,cm and 20\,cm data from different telescopes, the 20\,cm data were taken from the VLA-D
array and the 6\,cm data from the Green Bank 91m (7), the Parkes 64m (6), or the MERLIN array
(1). All data with the corresponding telescopes and, if necessary, the configuration, are
presented in Table\,2. Below, these data are used determine the nuclear
radio fluxes and corresponding luminosities.

\subsection{The nonmaser sample}

As mentioned before, the detection rate of extragalactic H$_{2}$O masers is rather low
(commonly $<$10\%, e.g., Henkel et al. 2005), and it is reasonable to ask whether H$_{2}$O maser
galaxies have ``special'' intrinsic properties, relative to apparently similar galaxies without
detected maser emission. Diamond-Stanic et al. (2009) have compiled a complete sample of 89 nearby
Seyfert galaxies, drawn from the unbiased revised Shapley-Ames catalog (with a limiting magnitude
of $B_{T}< 13$\,mag) (Maiolino \& Rieke 1995; Ho, Filippenko \& Sargent 1997). This sample is unique
in that it covers mostly Seyfert 2s (like our H$_2$O maser sample), that it is complete,
that the galaxies are rather nearby, and that sensitive H$_2$O measurements have been obtained
toward all these targets. The distance of all those sources is less than 200\,Mpc, which is comparable
to that of our H$_{2}$O maser galaxies (except for the two quasar sources, SDSS J0804+3607 with
$z$$\sim$0.66 and MG J0414+0534 with $z$$\sim$2.64, detected by Barvainis \& Antonucci 2005 and
Impellizzeri et al. 2008, respectively).

There are 19 maser and 70 nonmaser sources in the Seyfert comparison sample, the latter providing
a useful comparison. This nonmaser sample (its sources have been targeted in H$_{2}$O maser
surveys, but have not been detected) includes 16 Seyfert 1s (type 1-1.5) and 54 Seyfert 2s (type
1.8-2.0), which is similar to our maser-host Seyfert sample (most Seyfert 2s, just five Seyfert
1s (type 1-1.5), NGC\,2782, UGC\,5101, NGC\,235A, NGC\,4051 and NGC\,4151). Upper limits to their
maser luminosity were estimated from the RMS value (taken from the Hubble Constant Maser Experiment
(HoME\footnote{https://www.cfa.harvard.edu/$\sim$lincoln/demo/HoME}) and the Megamaser Cosmology
Project (MCP\footnote{https://safe.nrao.edu/wiki/bin/view/Main /MegamaserCosmologyProject
webpages}), assuming a characteristic linewidth of a spectral feature of 20\,km\,s$^{-1}$ and a
5$\sigma$ detection threshold (e.g., Bennert et al. 2009).

The 6\,cm and 20\,cm radio flux densities are collected from the literature. There are 61 sources
with 6\,cm data and 32 sources with 20\,cm data. For complementary 20\,cm data, the NVSS and FIRST
catalogs were also used, and we found additional 20\,cm data for 17 sources and upper limits for
12 sources (with relatively low angular resolution). The nonmaser sample and corresponding data
are presented in Table\,3.

\section{Analysis and discussion}

\subsection{Radio properties of H$_{2}$O maser galaxies}

\begin{figure}
\includegraphics[width=9cm]{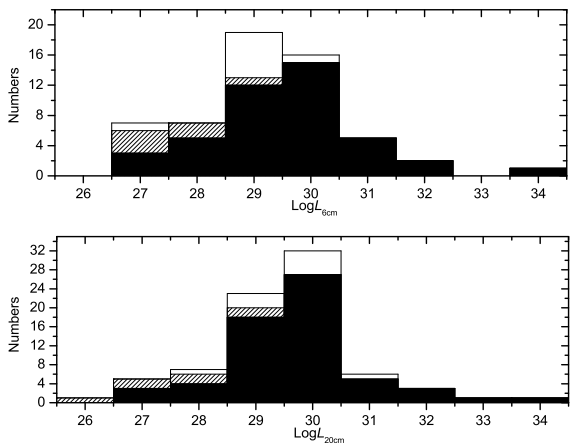}
\caption{The luminosity distributions of the 6\,cm (log$L_{\rm 6\,cm}$, upper panel) and
20\,cm (log$L_{\rm 20\,cm}$, lower panel) continuum emission in units of erg\,s$^{-1}$Hz$^{-1}$
for the entire maser sample, the AGN-maser subsample (shaded regions), the star formation (SF-)
maser subsample (diagonal lines), and masers of unknown type (white).}
\end{figure}

\begin{figure}
\includegraphics[width=8cm]{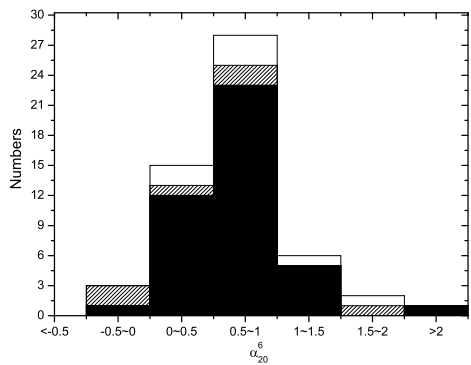}
\caption{Number distribution of the radio spectral index between 6\,cm and 20\,cm.
The shaded region of the histogram represents the subsample of AGN masers. The area with
diagonal lines denotes SF-masers, while the remaining masers are of unknown type.}
\end{figure}

For our H$_{2}$O maser galaxies with available data, the distributions of nuclear
radio luminosities at 6\,cm and 20\,cm are plotted in Figure 1. The upper panel
shows the histogram of 6\,cm luminosities for both the entire sample of 57 sources and
the subsample of 43 AGN-masers. For the entire sample, the distribution of
log$L_{\rm 6\,cm}$ peaks around 29 (here and throughout the article, $L_{\rm 6\,cm}$
and $L_{\rm 20\,cm}$ are given in units of erg\,s$^{-1}$\,Hz$^{-1}$) and the mean
value is 28.8$\pm$0.2 (Throughout the paper, given errors are the standard deviation
of the mean.) The mean values are 29.0$\pm$0.2, 27.2$\pm$0.4, and 28.5$\pm$0.3 for
AGN-masers, SF-masers, and masers without known type, respectively. The distribution
of the 20\,cm luminosity is presented in the lower panel, also for all of the 79
sources and for the subsample of 56 AGN-masers, respectively. The mean value of
log$L_{\rm 20\,cm}$ is 29.1$\pm$0.2 for the entire sample and the mean values are
29.3$\pm$0.2, 27.4$\pm$0.5, and 28.9$\pm$0.2 for AGN-masers, SF-masers, and masers
without known type, respectively. The subsample of AGN-disk-maser
galaxies (29 sources; not shown separately, but see section 3.3) shows
nuclear radio luminosities consistent with those of the entire AGN-sample.
Comparing the subsamples of AGN-masers and SF-masers, a striking difference can be found
between their luminosities at both 6\,cm and 20\,cm, i.e., AGN-masers are detected toward
galaxies with higher nuclear radio luminosities. This important point is discussed further
in section 3.3.

To study the radio properties of our H$_{2}$O maser galaxies, radio spectral indices
were derived for those sources for which 6\,cm and 20\,cm flux densities were
measured. Assuming a power-law dependence for the continuum flux density given by $S\propto
\nu^{-\alpha}$, the spectral index can be calculated by
 $$
 \alpha = \log(S_{6\,cm}/S_{20\,cm})/\log(6/20).
 $$


The spectral index ($\alpha$) is obtained for 55 sources including 42 targets hosting
AGN-masers, six objects exhibiting SF-masers, and seven sources with unknown maser type (see
the second last column of Table\,2). Figure 2 plots the distribution of the spectral
index for those 55 maser sources and the subsamples of AGN- and SF-masers. The
spectral index for the majority of sources lies in the 0.5$-$1.0 bin, with a mean value of
$0.66\pm0.07$. Since H$_{2}$O maser emission is mostly detected in Seyfert 2 systems,
it agrees with the fact that Seyfert 2 galaxies are always steep spectrum sources.
If targets with spectral index $\alpha \geq 0.3$ are defined as steep spectrum
sources (Ho \& Ulvestad 2001), 80\% (44/55) of our maser sources belong to this category.
For the subsamples of AGN-masers and SF-masers, the mean spectral indices are $0.66\pm0.07$
and $0.61\pm0.27$, respectively. No significant difference between both distributions
can be found. However, in view of the small number of SF-masers, statistical constraints
are not very stringent. In addition, we note that the spectral index {\it might} be
overestimated for those 12 sources with 6\,cm data from single-dish telescopes and 20\,cm
data from the VLA-D array. Eliminating these sources, the mean spectral index for the
remaining 32 AGN-maser sources is $0.65\pm0.07$, while the three remaining SF sources do
not justify a statistical evaluation. For the 11 sources with ``optimal'' data (9
AGN-masers and 2 targets of unknown type observed at 6\,cm with the VLA-B and at
20\,cm with the VLA A-array), the continuum appears to be flatter with a mean index
value of $0.38\pm0.12$.

\subsection{\textbf{Comparison of radio properties of maser and nonmaser galaxies}}


For the entire nonmaser sample (see Sect.\,2.2; all targets are Seyfert galaxies),
the mean log$L_{6\,cm}$ and log$L_{\rm 20\,cm}$ luminosities (in units of
erg\,s$^{-1}$\,Hz$^{-1}$) are 27.7 $\pm$0.2 and 28.2$\pm$0.2, respectively. For all
our maser Seyferts, the mean values are log$L_{\rm 6\,cm} = 28.8 \pm 0.2$ and log$L_{\rm
20\,cm} = 29.1 \pm 0.1$, respectively (see Table~1). The difference in nuclear
radio luminosity between nonmaser and maser Seyferts is obvious. Considering that
extragalactic H$_2$O masers are mostly found in Seyfert 2s, we can further compare the
nuclear radio luminosities between the nonmaser Seyfert 2 subsample (36 sources)
and our maser Seyfert 2 sources (41 sources). The mean values of log$\,L_{\rm 6\,cm}$ and
log$\,L_{20\,cm}$ (again in units of erg\,s$^{-1}$\,Hz$^{-1}$) are 27.5$\pm$0.2
and 28.0$\pm$0.2 for nonmaser Seyfert 2s and 28.8$\pm$0.2 and 29.1$\pm$0.2 for maser
Seyfert 2s, respectively.  It shows the same trend, i.e., maser Seyfert 2s have higher
nuclear radio luminosities than the nonmaser Seyfert 2s. The difference seems
to amount to slightly more than an order of magnitude, roughly corresponding to
factors of $\sim$10--20.

\begin{figure}
\centering
\includegraphics[width=9cm]{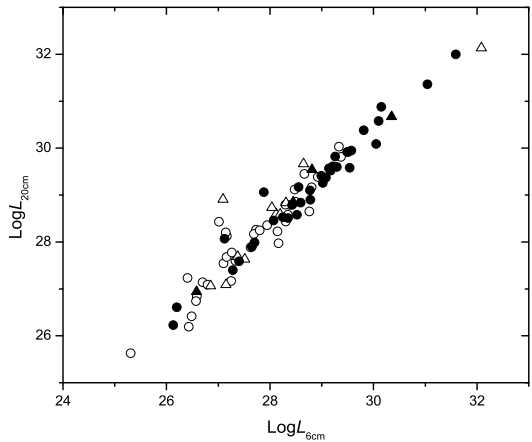}
\caption{Distribution of the 6\,cm against the 20\,cm luminosity of both the maser (filled
symbols) and the nonmaser Seyfert sample (open symbols). Circles and triangles show Seyfert
2s and Seyfert 1s, respectively. Units for $L_{\rm 6cm}$ and $L_{\rm 20cm}$ are
erg\,s$^{-1}$\,Hz$^{-1}$}.
\end{figure}

To visualize the difference, we plot 6\,cm versus 20\,cm luminosities for both the
nonmaser Seyfert sample and our maser Seyfert sample (Fig.\,3). The trend is apparent. Most
nonmaser sources are distributed in the lower left, while maser
galaxies are mostly located in the central and upper right regions.
However, there is also overlap, and there are a few maser Seyferts with low nuclear
radio luminosities. This indicates that there is no strict radio continuum luminosity
threshold for individual sources.

So far, we have not yet discussed a potential distance bias for the samples considered
here. If measurements are very sensitive and signals strong, different distances should
not play a major role with respect to detection rates. If one is, however, near the limit of
sensitivity, typical luminosities of observed and classified sources should increase with
the distance squared. While the average distances of the maser and nonmaser Seyfert samples
show a similar range (the two distant quasars detected in H$_2$O by Barvainis \& Antonucci
2005 and Impellizzeri et al. 2008 do not belong to any of the Seyfert samples), we find
for the mean values 66.4 $\pm$ 7.5 and 31.5 $\pm$ 2.6\,Mpc, respectively. Therefore, this
may introduce a bias of order $(66.4/31.5)^2 \sim 4$. Even fully accounting for this effect,
however, a luminosity ratio of order four still remains between the maser and the nonmaser samples.

In the nonmaser sample, there is just one source with distance $>$71\,Mpc, while there are 20
such sources in our maser Seyfert sample. Eliminating all sources with distances $>$71\,Mpc,
the mean distances of the maser and nonmaser sample become similar ($31.9\pm3.4$\,Mpc versus
$29.9\pm2.1$\,Mpc). The mean luminosities are then log$L_{\rm 6\,cm} = 28.4 \pm 0.3$ and
log$L_{20\,cm} = 28.7 \pm 0.2$ for our maser sample and log$L_{\rm 6\,cm} = 27.7 \pm 0.2$ and
log$L_{\rm 20\,cm} = 27.9 \pm 0.2$ for the nonmaser sample (units: erg\,s$^{-1}$\,Hz$^{-1}$).
The difference in their nuclear radio luminosities is still apparent (ratios of about
5 and 6 at 6\,cm and 20\,cm, respectively).

The luminosity ratio between the 16 H$_{2}$O detected galaxies (three additional sources
show SF-related H$_2$O or H$_2$O emission of unknown type) and the 70 H$_{2}$O undetected
sources in the entire Seyfert comparison sample is not affected by the Malmquist bias. Both
subsamples have similar mean distances ($27.5\pm7.7$ and $31.5\pm2.6$\,Mpc, respectively).
The mean nuclear radio luminosity of the 16 sources comprising the maser
subsample (log$L_{\rm 6\,cm} = 28.2 \pm 0.3$ and log$L_{\rm 20\,cm} = 28.4 \pm 0.2$) is about two
to three times higher than that of the 70 non-detections. The Seyfert 2 sources among these two
subsamples include 15 AGN-related H$_{2}$O detections and 54 non-detections with mean
distances of $28.5\pm7.4$\,Mpc and $29.8\pm2.3$\,Mpc, respectively. The mean luminosities on
an erg\,s$^{-1}$\,Hz$^{-1}$ scale are log$\,L_{\rm 6\,cm}$ = 28.2$\pm$0.3 and log$\,L_{\rm 20\,cm}$
= 28.5$\pm$0.3 for the first and log$\,L_{\rm 6\,cm}$ = 27.5$\pm$0.2 and log$\,L_{\rm 20\,cm}$ =
28.0$\pm$0.2 for the second sample. The luminosity ratio is of order four (five at 6\,cm and
three at 20\,cm). With all this evidence we thus conclude that a nuclear radio
continuum luminosity ratio greater than unity is real. For a summary, see Table~1.

\begin{table}
\caption[]{A comparison of nuclear radio luminosities of maser and nonmaser galaxies
(for details, see Sect.\,3.2)
\label{comparison}}
\begin{scriptsize}
\begin{center}
\begin{tabular}{c l c c}
\hline
Samples         & subsamples  &$Log\,L_{6\,cm}$ & $Log\,L_{20\,cm}$        \\
                &             &\multicolumn{2}{c}{(erg\,s$^{-1}$Hz$^{-1}$)}\\
\hline
Maser Seyferts  & Total       &  28.8$\pm$0.2   & 29.1$\pm$0.1             \\
                & Seyfert 2   &  28.8$\pm$0.2   & 29.1$\pm$0.2             \\
                & $<$71Mpc    &  28.4$\pm$0.3   & 28.7$\pm$0.2             \\

nonmasers      & Total       &  27.7$\pm$0.2   & 28.2$\pm$0.2             \\
(70 sources)    & Seyfert 2   &  27.5$\pm$0.2   & 28.0$\pm$0.2             \\
                & $<$71Mpc    &  27.7$\pm$0.2   & 27.9$\pm$0.2             \\
\hline
Non-Masers (70) & Total       &  27.7$\pm$0.2   & 28.2$\pm$0.2             \\
                & Seyfert 2   &  27.5$\pm$0.2   & 28.0$\pm$0.2             \\
Masers (19)     & Total       &  28.2$\pm$0.3   & 28.4$\pm$0.2             \\
                & Seyfert 2   &  28.2$\pm$0.3   & 28.5$\pm$0.2             \\
\hline
\end{tabular}
\end{center}
\end{scriptsize}
\end{table}

\begin{figure}
\centering
\includegraphics[width=10cm]{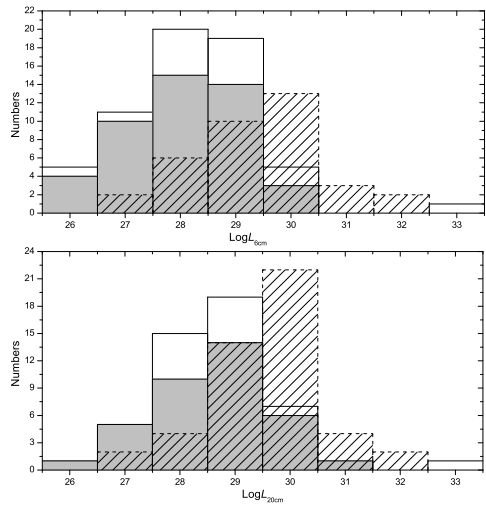}
\caption{Distribution of the 6\,cm (log$L_{\rm 6\,cm}$, upper panel) and 20\,cm luminosities
(log$L_{\rm 20\,cm}$, lower panel) in units of erg\,s$^{-1}$Hz$^{-1}$ for the sample of nearby
Seyfert galaxies without detected maser emission and its subsample of Seyfert 2s (shaded
regions). For comparison, the distributions for our maser Seyfert 2s are also presented
(areas filled with diagonal lines).}
\end{figure}

To visualize this result, 6\,cm and 20\,cm luminosity distributions are plotted for the
entire nonmaser sample and its subsample comprised of Seyfert 2s in Figure 4. For
comparison, the luminosity distributions for our maser Seyfert 2s are also presented.
The differences are significant for the two Seyfert 2 samples and a Kolmogorov-Smirnov
(KS) test shows probabilities of 0.005\% at 6\,cm and 0.01\% at 20\,cm that both Seyfert 2
samples are drawn from the same parent population.

\begin{figure}
\centering
\includegraphics[width=9cm]{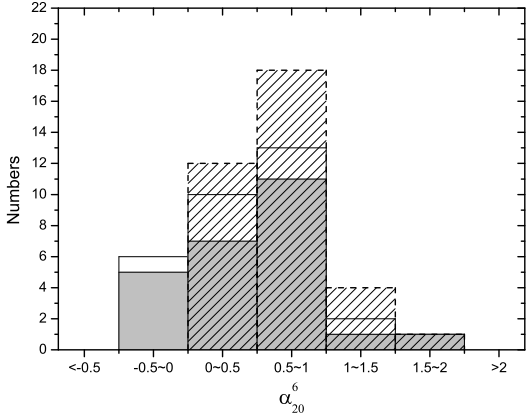}
\caption{The distributions of the spectral index between 6\,cm and 20\,cm for the nearby Seyfert
sample without detected maser emission and its subsample of Seyfert 2s (shaded regions). For comparison,
the distribution of Seyfert 2s with maser detections is also presented (areas filled with diagonal lines).}
\end{figure}

Similar to our maser sample, spectral indices were calculated for those 48 sources of the
comparison sample (including 16 sources with 20\,cm data from the NVSS or FIRST) with
both 6\,cm and 20\,cm flux density data. The mean values of the index are 0.84$\pm$0.12
and 0.83$\pm$0.12 for the entire sample (48 sources) and for the Seyfert 2 subsample (36
sources). However, as mentioned before, 16 sources have low-resolution 20\,cm data (from
NVSS or FIRST) and high-resolution 6\,cm data. This results in overestimated spectral indices
for these sources, even including extremely high values (e.g., 3.47 for NGC\,1097 and 2.89
for NGC\,2992). After excluding those 16 sources, the mean values become
0.52$\pm$0.09 for the entire sample (32 sources) and 0.55$\pm$0.10  for the Seyfert 2
subsample (24 sources). Figure 5 plots the distribution of the spectral index
for the nonmaser Seyfert sample (without including those 16 sources). Compared with maser host galaxies (dashed lines in Fig.\,5), no significant
difference can be found between the spectral index distributions. A KS test results in a
probability of 79.2\% that maser and nonmaser Seyfert 2 samples are drawn from the same
parent distribution.

\begin{figure}
\centering
\includegraphics[width=8.5cm]{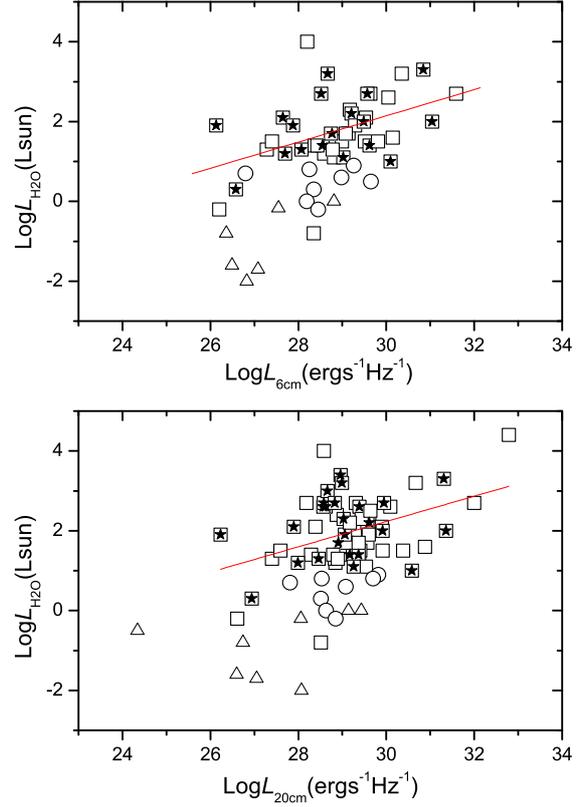}
\caption{H$_{2}$O maser luminosities  (logarithmic scale, in $L_{\odot}$) against the nuclear radio
luminosity (logarithmic scale, in erg\,s$^{-1}$Hz$^{-1}$) of maser host galaxies at 6\,cm (upper
panel) and 20\,cm (lower panel). Squares, triangles, and circles represent AGN-masers, star
formation-masers, and masers of unknown type, respectively. Pentacles inside squares indicate potential
or confirmed disk-maser sources, which are presented in boldface in Table\,2. The solid lines show linear
fits to the AGN-maser sources.}
\end{figure}

\subsection{H$_{2}$O maser power v.s. nuclear radio power}

Here we study possible correlations between H$_{2}$O maser line strength and AGN power.
For the SF-masers alone we do not expect a correlation between the nuclear radio
emission and maser power, since these maser spots are found in star-forming regions
well displaced from the nuclei.

Maser luminosities, assuming here and elsewhere isotropic emission, and nuclear radio
continuum fluxes are given in Table\,2. In Fig.\,6, these maser luminosities are plotted against
6\,cm (upper panel) and 20\,cm continuum luminosities (lower panel). For both panels, a correlation
can be found, i.e., rising maser luminosity with increasing nuclear radio luminosity.
SF-maser sources ($L_{\rm H_{2}O}<10L_{\odot}$) tend to be located in the lower left and AGN-masers
(mostly $L_{\rm H_{2}O}>10L_{\odot}$) in the upper right hand corners of Fig.\,6, i.e., AGN-maser galaxies
tend to have higher nuclear radio luminosities relative to SF-masers (see also Sect.\,3.2).

For the 6\,cm continuum and 1.3\,cm AGN-maser line luminosity (upper panel of Fig.\,6, 41 sources),
a weak correlation is apparent. A linear least-square fit shows log$L_{\rm H_2O}$ = (--7.7$\pm$3.1)
+ (0.3$\pm$0.1)log$L_{\rm 6\,cm}$, with Spearman's rank correlation coefficient $R$ = 0.51 and a chance
probability $P$ = 5.1$\times$10$^{-4}$. For the 20\,cm continuum and maser line luminosities (lower
panel, 61 sources), our linear fitting gives log$L_{\rm H_2O}$ = (--7.3$\pm$2.7)+(0.3$\pm$0.1)log\,$L_{\rm
20\,cm}$, with the fitting parameters $R$ = 0.41 and $P$ = 1.0$\times$10$^{-3}$. As already mentioned,
for many sources, the 6\,cm and 20\,cm data were obtained with different beam sizes (for details,
see Table\,2). It is thus necessary to consider this effect on the correlation. For the nine AGN-masers
among those 11 sources with `ideal' observational data (i.e., 6\,cm data from the VLA-B and 20\,cm
data from the VLA-A configuration), the correlation becomes log$L_{\rm H_2O}$=(--8.1$\pm$6.0) +
(0.3$\pm$0.2)$log\,L_{\rm 6\,cm}$, with the fitting parameters $R$=0.51 and $P$=0.16. While they
are too few sources for a reliable fit, the tendency toward higher continuum flux densities
associated with more luminous masers is still apparent.

\begin{figure}
\centering
\includegraphics[width=9cm]{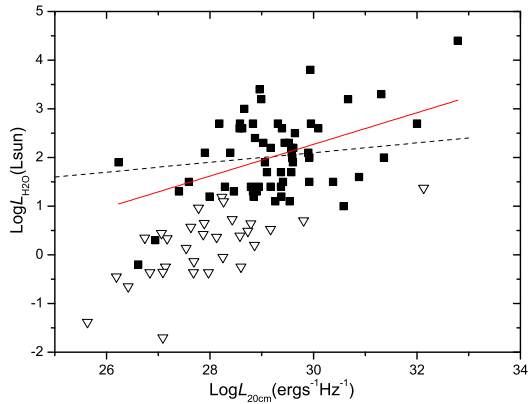}
\caption{H$_{2}$O maser luminosity (logarithmic scale, in $L_{\odot}$) versus 20\,cm luminosity
(logarithmic scale in erg\,s$^{-1}$Hz$^{-1}$) for AGN-masers (squares). The solid (red)
and the dashed (black) lines show a linear fit without and with considering the Malmquist
effect, respectively. The non-detection sample is also plotted displaying the upper limits of the
maser luminosity (triangles).}
\end{figure}

\begin{figure}
\centering
\includegraphics[width=9cm]{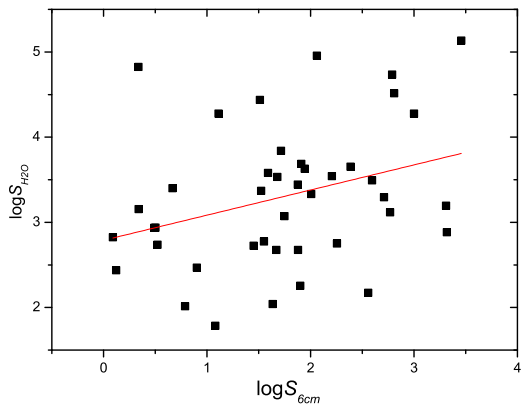}
\caption{Integrated H$_{2}$O flux densities (logarithmic scale, in Jy\,km\,s$^{-1}$) versus 6\,cm
luminosity (logarithmic scale in Jy) for AGN-masers (squares). The red line shows a very tentative
linear fit.}
\end{figure}

As a subsample of the AGN maser sources, disk maser candidates are also shown in Figure 6. These masers originate in the innermost few parsec, which might suggest a particularly close connection between their maser emission and the AGN-power. A correlation seems
again to be apparent ($R$=0.37 and $P$=0.13), but the small number of disk-maser sources mean the
statistical evidence is still quite weak.

We have not yet considered the Malmquist bias caused by different mean distances of galaxies
comprising the samples for the luminosity-luminosity correlations. Thus we reanalyzed the relation between
nuclear radio continuum and maser luminosities, using the method of partial correlation
coefficients (e.g., Darling \& Giovanelli 2002; Kandalyan \& Al-Zyout 2010). Since both luminosities
are correlated with the variable distance, the correlations should be subtracted when we analyze the
correlation between two luminosities.  This type of correlation is known as a partial correlation
coefficient. Figure 7 plots H$_{2}$O maser luminosity against the 20\,cm nuclear radio
luminosity for our AGN-maser subsample, and shows linear fits without and with considering the Malmquist effect. The correlation
becomes weaker after taking the Malmquist effect into account. The correlation coefficient changes
from $\sim$0.3 to $\sim$0.2 and the fitting becomes flatter with a slope of $\sim$0.2 instead of $\sim$0.3.

For comparison, our non-detection sample with upper limits to the maser luminosity is also presented
in Fig.\,7. The upper limits were obtained from the individual RMS values of H$_{2}$O maser data (see
details in Sect.\,3.2 and Table 3). A few overlaps in maser luminosity can be found, along with H$_2$O undetected
sources located in the lower left hand region well below the fitting line. This is consistent with
the trend described in Sect.\,3.2, i.e., maser activity tends to weaken with decreasing nuclear radio power.

For the AGN-maser subsample, we also correlated integrated line and continuum flux densities
(for the 6\,cm continuum, see Fig.\,8). A linear fit yields log($S_{\rm H_2O}$$\times$$\Delta V$) =
(0.3$\pm$0.1)log\,$S_{\rm 6\,cm}$ + (2.8$\pm$0.3) (41 sources) and log($S_{\rm H_2O}$$\times$$\Delta V$)
= (0.2$\pm$0.1)log$\,S_{\rm 20\,cm}$ + (3.0$\pm$0.2) (56 sources), respectively, with Spearman's rank
coefficients amounting to $R$ = 0.32 and 0.19 and chance probabilities of $P$ = 0.03 and 0.16.

There may be a variety of causes for our statistically marginal correlations between maser
 and nuclear radio continuum. In the case of unsaturated H$_2$O maser emission only a part of the
nuclear continuum may be amplified (see, e.g., the situation in NGC\,4258, Herrnstein et al. 1998). For saturated maser emission, no direct correlation with the background continuum will appear. In
addition, the AGN-maser sample presented here is inhomogeneous. It includes disk masers and jet masers
(masers related to the interaction of nuclear radio jets with ambient molecular clouds or to the
amplification of the jet's seed photons by suitably located foreground clouds), and perhaps also
additional so far not yet identified maser types. The absorbing column densities along the lines of
sight also differ. Among 31 AGN-masers with available data, about 60\% are Compton-thick (Zhang et al.
2010). Moreover, 6\,cm or 20\,cm measurements as indicators of intrinsic nuclear power may introduce
some uncertainties. The observed nuclear radio luminosity may depend on the geometry of the
system, such as shielding by the torus. And large uncertainties may be produced from isotropically
calculated H$_{2}$O maser luminosities, since maser beam angles are often poorly constrained and
may be quite small due to source geometry and propagation effects (Kartje, K{\"o}nigl \& Elitzur
1999). Lastly, uncertainties in the nuclear radio luminosity can be introduced
from radio data including measurements with different resolution and sensitivity. In view of all
these uncertainties, the presence of a correlation would hint at the possibility that a higher
degree of activity may be able to heat up a larger volume to temperatures suitable for 22\,GHz
H$_2$O maser emission (Neufeld et al. 1994). More detailed interferometric studies are therefore
required to obtain a larger sample of ``ideal'' data for deeper insight.

\section{Summary}

In this paper, the entire extragalactic 22\,GHz H$_{2}$O maser sample published so far has been
studied, with the notable exception of the Magellanic Clouds. Radio continuum data were
analyzed to better understand the properties of the AGN associated with H$_{2}$O
masers. For comparison, a complementary Seyfert sample without detected maser emission was also
compiled. The main results are summarized in the following.

(1) For the entire H$_{2}$O Seyfert maser sample, the mean 6\,cm  and 20\,cm nuclear radio
luminosities are log$\,L_{\rm 6\,cm}$=28.8$\pm$0.2 and log$\,L_{\rm 20\,cm}$=29.1$\pm$0.2, respectively
($L_{6\,cm}$, $L_{20\,cm}$ in units of erg\,s$^{-1}$Hz$^{-1}$). For the comparison sample of Seyferts
without detected maser emission, the mean values of log$L_{6\,cm}$ and log$L_{20\,cm}$ are 27.7$\pm$0.2
and 28.2$\pm$0.2, respectively. Comparisons of corresponding subsamples also show the difference,
i.e., H$_{2}$O maser sources tend to be more luminous than nonmaser sources. After
considering a potential distance bias, we converged on a luminosity ratio of order five.

(2) For the H$_{2}$O maser sample and the sample of Seyferts devoid of detected H$_{2}$O emission,
spectral indices were derived from the 6\,cm and 20\,cm data. There is no significant difference
for the distributions of the spectral indices between our H$_{2}$O maser sources and the H$_{2}$O
undetected sources. The mean value of the index is $0.66\pm0.07$ for our H$_{2}$O maser sample
and $0.52\pm0.09$ for the comparison sample.

(3) For the subsample of AGN-masers, a correlation seems to be present between H$_{2}$O maser
isotropic luminosity and nuclear radio power of the host galaxy. However, the correlation is
significantly affected by the Malmquist bias so that no definite conclusion can be drawn right now.

(4) Based on the results outlined above, the nuclear radio luminosity may provide a clear signature
 of AGN-masers, possibly providing suitable constraints for future H$_{2}$O megamaser surveys.

\acknowledgements
This work is supported by the Natural Science Foundation of China (No. 11043012, 11178009) and
China Ministry of Science and Technology under State Key Development Program for Basic Research (2012CB821800). We made use
of the NASA Astrophysics Data System Bibliographic Services (ADS) and the NASA/IPAC extragalactic
Database (NED), which is operated by the Jet Propulsion Laboratory, California Institute of
Technology, under contract with NASA.

\clearpage

\onecolumn
\scriptsize
\begin{longtable}{lllllllll}
 \caption{Radio properties of H$_{2}$O maser host galaxies$^{*}$}\\
\hline
 Source        &  Type    &  D  &  S$_{6}$   & Tel./Ref.   & S$_{20}$ & Tel./Ref. &  $\alpha$ &  log\,$L_{\rm H_{2}O}$         \\

\hline
\endfirsthead
\caption{continued.}\\
\hline
 Source        &  Type    & D   &  S$_{6}$   & Tel./Ref.   & S$_{20}$ & Tel./Ref. &  $\alpha$ &  log\,$L_{\rm H_{2}O}$         \\
\hline
\endhead
\hline
\endfoot

\textbf{NGC\,17}     &Sy2          &79.1  &              &             &$67.5\pm2.5$     & VLA-D/Con98    &        &     0.8   \\
 NGC\,23             &LINER        &60.9  &$33.0\pm5$    & GB/Bec91    &$74.3\pm2.7$     & VLA-D/Con98    &  0.67  &     2.3   \\
\emph{IC\,10}        &             &1.2   &$137.0\pm14$  & GB/Gre91    & 319.0           & GB/Whi92       &  0.70  &    -0.8   \\
NGC\,0235A           & Sy1         &86.90 &              &             &$42.6\pm1.4$     & VLA-D/Con98    &        &     2.0   \\
NGC\,253             & Sy2,SBG     &3     & 2080.0       &Parkes/Wri90 &$2994.7\pm113.5$ & VLA-D/Con98    &  0.30  &    -0.8   \\
NGC\,262 (Mrk\,348)  & Sy2         &62.00 &$244.0\pm3.7$ &GB/Gre91     &$292.7\pm8.8$    & VLA-D/Con98    &  0.08  &     2.6   \\
IRAS F01063-8034     &             &67    &$81.0\pm7$    &Parkes/Wri94 &                 &                &        &     2.7   \\
NGC\,449 (Mrk\,1)    & Sy2         &64    &$28.0\pm2$    &VLA-A/Ulv84a &$75.9\pm2.3$     & VLA-D/Con98    &  0.83  &     1.7   \\
NGC\,520             & SBG         &27    &$90.0\pm14$   &GB/Bec91     &  182.0          & VLA-D/Con98    &  0.28  &     0.3   \\
\emph{NGC\,598(M\,33)}& H\,II      &0.7   &              &             &$3.7\pm0.6$      & VLA-D/Con98    &        &    -0.5   \\
\textbf{NGC\,591}    &Sy2          &61    &  7.9         &VLA-A/Ulv89  &24.5             & VLA-A/Ulv89    &  0.9   &     1.4   \\
NGC\,0613            &Sy2          &17.9  &$101.0\pm13$  &Parkes/Wri96 &179.6            & VLA-D/Con98    &  0.48  &     1.2   \\
IC\,0184             &Sy2          &70.5  &              &             &$3.3\pm0.5$      & VLA-D/Con98    &        &     1.4   \\
NGC\,1052            &Sy2,LINER    &17    &1000.0        &VLA-A/Bra97  &1103.0           & VLA-A/Coo07    &  0.08  &     2.1   \\
\textbf{NGC\,1068}   &Sy2,Sy1      &14.5  & 644.0 &VLA-B /Ho01         &1630.0           & VLA-A/Ho01     &  0.77  &     2.2   \\
NGC\,1106            &Sy2          &57.8  &$46.0\pm6.9$  &GB/Gre91     &166.0            & GB/Whi92       &  1.07  &     0.9   \\
Mrk\,1066            &Sy2          &48    &  35.5        &VLA-A/Ulv89  &94.3             & VLA-A/Ulv89    &  0.81  &     1.5   \\
\textbf{NGC\,1320}   &Sy2          &35.5  &   3.3        &VLA-D/Gal06  &$6.5\pm0.6$      & VLA-D/Con98    &  0.56  &     1.2   \\
\textbf{NGC\,1386}   &Sy2          &17    &  13.0        &VLA-A/Ulv84b &23.0             & VLA-A/Ulv84b   &  0.47  &     2.1   \\
IRAS 03355+0104      &Sy2          &159.1 &              &             &$26.5\pm1.3$     & VLA-D/Con98    &        &     2.1   \\
\emph{IC\,342}       &Sy2          &2     &$277.0\pm41.5$&GB/Gre91     &2475.0           & GB/Whi92       &  1.82  &    -2.0   \\
MG J0414+0534        &QSO1         &10836 &$1119.0\pm167$&GB/Bec91     &2676.0           & GB/Whi92       &  0.72  &     4.0   \\
\textbf{UGC\,3193}   &             &59.4  &              &             &$17.7\pm0.7$     & VLA-D/Con98    &        &     2.4   \\
UGC\,3255            &Sy2          &75    &              &             &$35.8\pm1.2$     & VLA-D/Con98    &        &     1.2   \\
Mrk\,3               &Sy2          &54    &$361.0\pm30$  &VLA-A/Wil80  &$1100.9\pm33$    & VLA-D/Con98    &  0.93  &     1.0   \\
\emph{NGC\,2146}     &H\,II        &14.5  &              &             &$1074.5\pm40$    & VLA-D/Con98    &        &     0.0   \\
VII\, ZW 073         &Sy2          &158.9 &              &             &$12.4\pm0.6$     & VLA-D/Con98    &        &     2.2   \\
NGC\,2273            &Sy2          &24.5  &  24.5        &VLA-B/Ho01   &47.3             & VLA-A/Ho01     &  0.55  &     0.8   \\
\textbf{UGC\,3789}   &Sy2          &44.3  &              &             &$17.6\pm1.0$     & VLA-D/Con98    &        &     2.6   \\
Mrk\,78              &Sy2          &150   &$12.0\pm2$    &VLA-A/Ulv84a &$31.0\pm3$       & VLA-A/Ulv84a   &  0.79  &     1.5   \\
J0804+3607           &QSO2         & 2640 &              &             &73.7             &VLA-B/Bec95     &        &     4.4   \\
\emph{He 2-10}       &SBG          & 10.5 &$27.0\pm0.11$  &VLA-B/Zij90 &$21.1\pm1.2$     &VLA-BnA/Kob99   & -0.2   &    -0.2   \\
\textbf{2MASX J08362280}&Sy2       & 197.4&              &             &2.0              &VLA-B/Bec95     &        &     3.4   \\
Mrk\,1210            &Sy2,Sy1      &  54  &$56.0\pm10$   &GB/Bec91     &114.0            &VLA-D/Con02     &  0.59  &     1.9   \\
\textbf{NGC\,2639}   &LINER        &  44  &  182.0       &VLA-B/Ho01   & 102             &VLA-A/Ho01      & -0.48  &     1.4   \\
NGC\,2782            &Sy1,SBG      &  34  &$47.0\pm7.0$  &GB/Bec91     &252.0            &GB/Whi92        &  1.39  &     1.1   \\
NGC\,2824 (Mrk\,394) &Sy?          &  37  &              &             &$9.3\pm0.5$      &VLA-D/Con98     &        &     2.7   \\
SBS\,0927+493        &LINER        & 135.6&              &             &$9.3\pm0.5$      &VLA-D/Con98     &        &     2.7   \\
\textbf{NGC\,2960}   &LINER        &  66  &              &             &7.1              &VLA-D/Con02     &        &     2.6   \\
UGC\,5101            &LINER,Sy1.5  & 157.4&$76.0\pm11.4$ &GB/Bec91     &158.0            &GB/Whi92        &  0.61  &     3.2   \\
NGC\,2979            &Sy2          &  36  &              &             &$15.7\pm1.0$     &VLA-D/Con02     &        &     2.1   \\
NGC\,2989            &H\,II        &  55.3&              &             &$18.6\pm1.5$     &VLA-D/Con02     &        &     1.6   \\
\emph{NGC\,3034(M\,82)}&SBG        &  3.7 &$3957.0\pm59$ &GB/Bec91     &8363.0           &GB/Whi92        &  0.62  &     0.0   \\
\textbf{NGC\,3079}   &Sy2,LINER    &  15.5& 114.0        &VLA-B/Ho01   & 132.0           &VLA-A/Ho01      &  0.12  &     2.7   \\
\textbf{Mrk\,34}     &Sy2          &  205 &   6.1        &VLA-A/Ulv84a &16.5             &VLA-A/Ulv84a    &  0.83  &     2.0   \\
\emph{NGC\,3359}     &H\,II        &  13.5&              &             &52.9             &VLA-D/Con02     &        &    -0.2   \\
\textbf{IC\,2560}    &Sy2          &  35  &              &             &31.0             &NVSS            &        &     3.0   \\
\textbf{NGC\,3393}   &Sy2          &  50  &$52\pm11$     &Parkes/Gri94 &$81.5\pm3.3$     &VLA-D/Con98     &  0.37  &     2.6   \\
NGC\,3556            &H\,II        &  12  &$90.0\pm13$   &GB/Gre91     &245.0            &GB/Whi92        &  0.83  &     0.0   \\
Arp\,299 (NGC\,3690) &SBG          & 42   &11.1$\pm$0.6  &VLA-A/Neff04 &41.6$\pm$2.1     & VLA-A/Neff04   & 1.1    &     1.4   \\
NGC\,3735            &Sy2          &  36  &   1.24       &VLA-B/Ho01   &1.61             &VLA-A/Ho01      &  0.22  &     1.3   \\
\emph{Antennae}      &SBG          &  20  &              &             &                 &                &        &     0.9   \\
\textbf{NGC\,4051}   &Sy1.5        &  10  &  3.2         &VLA-B/Ho01   &7.4              &VLA-A/Ho01      &  0.69  &     0.3   \\
NGC\,4151            &Sy1.5        &  13.5&  129.0       &VLA-B/Ho01   &324.0            &VLA-A/Ho01      &  0.76  &    -0.2   \\
\emph{NGC\,4214}     &SBG          &  2.94&$30.0\pm7$    &GB/Bec91     & 38.3            &VLA-D/Con02     &  0.2   &    -1.6   \\
\textbf{NGC\,4258}   &Sy1.9,LINER  &  7.2 &   2.2        &VLA-B/Ho01   & 2.7             &VLA-A/Ho01      &  0.19  &   1.9     \\
NGC\,4293            &LINER        &  17  &   1.8        &MERLIN/Fil06 & 18.5            &VLA-D/Con02     &  1.93  &   0.7     \\
\textbf{NGC\,4388}   &Sy2          &  34  &   26.9       &VLA-B/Ho01   &43.3             &VLA-A/Ho01      &  0.40  &   1.1     \\
NGC\,4527            &LINER        &  23.1&$151\pm22.5$  &GB/Gre91     &187.9            &VLA-D/Con02     &  0.18  &   0.6     \\
\textbf{ESO\,269-G012}&Sy2         &  66  &              &             &$<$3.7           &ATCA/Oos07      &        &   3.0     \\
NGC\,4922            &Sy2,LINER    &  95  &              &             & 40.1            &FIRST           &        &   2.3     \\
\textbf{NGC\,4945}   &Sy2          &   4  & 2840         &Parkes/Wri90 &6600             &Parkes/Wri90    &  0.26  &   1.7     \\
NGC\,5194 (M\,51)    &Sy2          &10    &  1.3         &VLA-B/Ho01   & 3.4             &VLA-A/Ho01      &  0.8   &   -0.2    \\
\emph{NGC\,5253}     &SBG          &3.33  &$90\pm12$     &Parkes/Wri96 & 83.8            &VLA-D/Con96     &  -0.06 &   -1.7    \\
NGC\,5256 (Mrk\,266) &Sy2,SBG      &112   &$43.3\pm2.2$  &GB/Bec91     & 159.0           &GB/Gre91        &  1.08  &   1.5     \\
NGC\,5347            &Sy2          &31    &   2.2        &VLA-A/Ulv89  & 3.4             &VLA-A/Ulv89     &  0.36  &   1.5     \\
\textbf{NGC\,5495}   &Sy2          &87.9  &              &             & 11.5            &NVSS            &        &   2.3     \\
\textbf{Circinus}    &Sy2          &4     & 610.0        &Parkes/Wri90 & 1500.0          &Parkes/Wri90    &  0.75  &   1.3     \\
NGC\,5506 (Mrk1376)  &Sy1.9        &25    &$160.0\pm8.0$ &VLA-A/Ulv84b & 315.0           &VLA-A/Ulv84b    &  0.56  &   1.7     \\
NGC\,5643            &Sy2          &16    &$87.0\pm10$   &Parkes/Wri94 & 203.0           &VLA-D/Con96     &  0.7   &   1.4     \\
\textbf{NGC\,5728}   &Sy2          &37    &   4.6        &VLA-A/Ulv89  & 70.0            &NVSS            &  2.26  &   1.9     \\
\textbf{UGC09618NED02}&LINER       &134.6 &$39.0\pm6.0$  &GB/Bec91     & 81.6            &VLA-D/Con02     &  0.61  &   3.2     \\
\textbf{NGC\,5793}   &Sy2          &47    & 508.0        &VLA-A/Ban06  & 1047.0          &VLA-A/Ban06     &  0.6   &   2.0     \\
NGC\,6240            &Sy2          &98    &  80.0        &VLA-A/Bra97  & 426.0           &VLA-D/Con02     &  1.39  &   1.6     \\
\textbf{NGC\,6264}   &Sy2          &135.7 &              &             & $<$0.9          &FIRST           &        &   3.1     \\
NGC\,6300            &Sy2          & 15   &$39.0\pm7.0$  &Parkes/Wri94 &                 &                &        &   0.5     \\
\textbf{NGC\,6323}   &Sy2          &104   &              &             & 3.1             &VLA-D/Con02     &        &   2.7     \\
ESO\,103-G035        &Sy2          & 53   &              &             &                 &                &        &   2.6     \\
IRAS F19370-0131     &Sy2          & 80   &              &             & 11.3            &NVSS            &        &   2.2     \\
\textbf{3C\,403}     &FR\,II       & 235  &$2026.0\pm283$&GB/Gre91     & 5798.0          &GB/Whi92        &  0.9   &   3.3     \\
\textbf{NGC\,6926}   &Sy2          & 80   &$48.0\pm11$   &Parkes/Gri95 & 117.0           &VLA-D/Con02     &  0.74  &   2.7     \\
AM2158-380NED02      &Sy2          & 128.8&  590.0       &Parkes/Wri90 & 1530.0          &Parkes/Wri90    &  0.79  &   2.7     \\
\textbf{TXS\,2226-184}&LINER       & 100  &$32.2\pm0.98$ &VLA-B/Tay02  &$73.3\pm2.2$     &VLA-B/Tay02     &  0.68  &   3.8     \\
NGC\,7479            &Sy2          & 31.8 &  3.1         &VLA-B/Ho01   &  4.0            &VLA-A/Ho01      &  0.21  &   1.3     \\
IC\,1481             &LINER        & 82   &              &             & 36.2            &VLA-D/Con02     &        &   2.5     \\

\end{longtable}

$^{*}$ Radio properties for the published extragalactic  H$_{2}$O maser host galaxies with luminosity distance $D$ $>$ 0.5\,Mpc (Zhang et al. 2010). Ten masers associated with star-forming regions are presented in italics and 29 disk-maser candidates in boldface.

 Column 1: Source; For Arp\,299, the emission from NGC\,3690 was only taken (Neff et al. 2004; Tarchi et al. 2011).

 Column 2: Types of nuclear activity from Zhang et al. (2010). SBG: starburst galaxy; Sy1.5, Sy1.9, Sy2: Seyfert type; LINER: low-ionization nuclear
           emission line region; ULIRG: ultra luminous infrared galaxy; FR\,II: Fanaroff-Riley type II radio galaxy; H\,II: H\,II galaxies, i.e., dwarf galaxies undergoing a burst of star formation (i.e., Bordalo et al. 2009);

 Column 3: The luminosity distance in Mpc, assuming $H_0$=75\,km\,s$^{-1}$\,Mpc$^{-1}$. For the high z objects (SDSS J0804+3607 and
           J0414+0534), $\Omega_{M}$ = 0.270 and $\Omega_{vac}$ = 0.730. 78 sources from Bennert et al. (2009), MG J0414+0534 from Impellizzeri et al. (2008), NGC\,17 and NGC\,1320 from Greenhill et al. (2008), He\,2-10, the Antennae, NGC\,4214 and NGC\,5253 from Darling et al. (2008);

 Columns 4 and 6: The 6\,cm and 20\,cm radio flux densities with uncertainties (if available) in mJy from ``Photometry \& SEDs'' in NED;

 Columns 5 and 7: Tel./Ref., used telescopes and corresponding reference for the 6\,cm and 20\,cm flux density. GB: Green Bank 91m telescope;
           VLA-A, -B, -C, -D: Very Large Array and used configuration; Parkes: the Parkes 64-m radio telescope; MERLIN: the multi-element radio-linked interferometer network; ATCA: Australian telescope compact array; NVSS: NRAO VLA sky survey catalog; FIRST: faint images of the radio sky at twenty-centimeters catalog.
           Ban06: Bann \& Kl{\"o}ckner (2006); Bec91: Becker et al. (1991); Bec95: Becker et al. (1995); Bra97: Braatz et al. (1997);
           Con83: Condon (1983); Con96: Condon et al. (1996); Con98: Condon et al. (1998); Con02: Condon et al. (2002);
           Coo07: Cooper et al. (2007); Fil06: Filho et al. (2006); Gal06: Gallimore et al. (2006); Gre91: Gregory \& Condon (1991);
           Gri95: Griffith et al. (1995); Ho01: Ho \& Ulvestad (2001); Kob99: Kobulnicky \& Johnson 1999; Neff04: Neff et al. (2004); Oos07: Oosterloo et al. (2007); Tay02: Taylor et al. (2002); Ulv84a: Ulvestad \& Wilson (1984a);  Ulv84b: Ulvestad \& Wilson (1984b); Ulv89: Ulvestad \& Wilson (1989); Whi92: White \& Becker (1992); Wil80: Wilson et al. (1980); Wri90: Wright \& Otrupcek (1990); Wri94: Wright et al. (1994); Wri96: Wright et al. (1996); Zij90: Zijlstra et al. (1990);

 Column 8: The spectral index between 6\,cm and 20\,cm, assuming $S \propto \nu^{-\alpha}$;

 Column 9: The apparent luminosity of maser emission (on a logarithmic scale), in units of $L_{\odot}$, taken from
           Bennert et al. (2009), Darling et al. (2008), Greenhill et al. (2008), and Tarchi et al. (2011).

\normalsize

\clearpage

\scriptsize
\begin{longtable}{lcccclccl}

 \caption{Radio properties of Seyfert galaxies without detected H$_{2}$O maser$^{*}$}\\
\hline
 Source        &  Type    &  D  &  S$_{6}$   & S$_{20}$    & Reference & $\alpha$  & RMS-H$_{2}$O  &  UL-H$_{2}$O      \\

\hline
\endfirsthead
\caption{continued.}\\
\hline
Source         &  Type    &  D  &  S$_{6}$   & S$_{20}$    & Reference & $\alpha$  & RMS-H$_{2}$O  &  UL-H$_{2}$O      \\
\hline
\endhead
\hline
\endfoot

 NGC\,777  &     2   &   66.5  &    2.63     &3.19        &Ho01            & 0.16  & 6.9  &1.2    \\
 NGC\,788  &     2   &   54.1  &    1.2      &2.2         &Ulv89           & 0.50  & 3    &0.6   \\
 NGC\,1058 &     2   &    9.2  &    $<$0.12  & 6.9        &Ho01, NVSS      &       & 20.0 &-0.1   \\
 NGC\,1097 &   1.0   &   16.5  &    3.8      & 249.2      &Mor99, NVSS     &3.47   & 15   &0.3  \\
 NGC\,1241 &     2   &   53.8  &    6.8      &            &The00           &       & 7.4  &1.0  \\
 NGC\,1275 &   1.5   &   70.1  &    20700    &23200       &Ho01            & 0.09  & 9.5  &1.4  \\
 NGC\,1365 &   1.8   &   21.5  &    0.9      &2.5         &San95           & 0.85  & 2.4  &-0.2  \\
 NGC\,1358 &     2   &   53.6  &    7.04     &17.9        &Ho01            & 0.78  & 3    &0.6  \\
 NGC\,1433 &     2   &   13.3  &    $<$3.0   &            &Sad95           &       & 15   &0.1   \\
 NGC\,1566 &   1.5   &   19.4  &    $<$6.0   &            &Sad95           &       & 12   &0.4  \\
 NGC\,1667 &     2   &   61.2  &    1.19     &4.08        &Ho01            & 1.02  & 6.6  &1.1  \\
 NGC\,2685 &     2   &   16.2  &    $<$0.13  &$<$1.0      &Ho01, FIRST     &       & 19.0 &0.4  \\
 NGC\,2655 &     2   &   24.4  &    44.1     &101         &Ho01            & 0.69  & 5.2  &0.2  \\
 NGC\,2992 &   1.9   &   34.1  &    7.0      & 226.2      &Sad95, NVSS     & 2.89  & 2.3  &0.1  \\
 NGC\,3031 &   1.5   &   3.6   &    91.2     &79.2        &Ho01            & -0.12 & 3.0  &-1.7   \\
 NGC\,3081 &     2   &   34.2  &    0.9      &2.5         &Ulv89           & 0.85  & 2.3  &0.1  \\
 NGC\,3147 &     2   &   40.9  &    10.1     &13.6        &Ho01            & 0.25  & 6.3  &0.7  \\
 NGC\,3185 &     2   &   21.3  &    0.47     &3.16        &Ho01, FIRST     &1.58   & 3    &-0.2  \\
 NGC\,3227 &   1.5   &   20.6  &    25.9     &78.2        &Ho01            & 0.92  & 2.6  &-0.2  \\
 NGC\,3254 &     2   &   23.6  &    $<$0.12  & $<$0.84    &Ho01, FIRST     &       & 28   &0.9  \\
 NGC\,3281 &     2   &   44.7  &    26.7     &61.2        &Ulv89           & 0.69  & 3.3  &0.5  \\
 NGC\,3486 &     2   &    7.4  &    $<$0.12  &$<$0.95     &Ho01, FIRST     &       & 21   &-0.2   \\
 NGC\,3516 &   1.2   &   38.9  &    6.03     &29.8        &Ho01            & 1.33  & 4    &0.5  \\
 IRAS11215 &     2   &   62.4  &    18       & 52.1       &Sch01, NVSS     & 0.88  & 20   &1.6  \\
 NGC\,3783 &   1.2   &   36.1  &    13       & 43.6       &Sad95, NVSS     & 1.01  & 5    &0.5  \\
 NGC\,3941 &     2   &   18.9  &    0.19     & $<$0.97    &Ho01, FIRST     &       & 22   &0.6  \\
 NGC\,3976 &     2   &   37.7  &    0.41     & $<$0.95    &Ho01, FIRST     &       & 16   &1.1  \\
 NGC\,3982 &   1.9   &   17.0  &    1.79     &3.56        &Ho01            & 0.57  & 3.0  &-0.4  \\
 NGC\,4138 &   1.9   &   17.0  &    0.78     &0.45        &Ho01            & -0.46 & 2.4  &-0.5  \\
 NGC\,4168 &   1.9   &   16.8  &    5.24     &4.37        &Ho01            & -0.15 & 15.0 &0.3  \\
 NGC\,4235 &   1.2   &   35.1  &    5.61     &5.05        &Ho01            & -0.09 & 4.2  &0.4  \\
 NGC\,4378 &     2   &   35.1  &    0.21     & $<$0.94    &Ho01, FIRST     &       & 15.0 &1.0  \\
 NGC\,4395 &   1.8   &   4.6   &    0.80     &1.68        &Ho01            & 0.62  & 3.8  &-1.4   \\
 NGC\,4472 &     2   &   16.8  &    65.2     &113         &Ho01            & 0.46  & 17.0 &0.4  \\
 NGC\,4477 &     2   &   16.8  &    0.18     & $<$0.97    &Ho01, FIRST     &       & 4.2  &-0.2  \\
 NGC\,4501 &     2   &   16.8  &    1.14     &2.06        &Ho01            & 0.49  & 3.0  &-0.4  \\
 NGC\,4507 &     2   &   59.6  &    10.8     & 66.1       &Bra98, NVSS     &1.50   & 9.8  &1.2  \\
 NGC\,4565 &   1.9   &    9.7  &    2.72     &2.31        &Ho01            & -0.14 & 4.6  &-0.7   \\
 NGC\,4579 &   1.9   &   16.8  &    43.1     &27.8        &Ho01            & -0.36 & 3.0  &-0.4  \\
 NGC\,4593 &   1.0   &   41.3  &    1.6      &2.1         &Ulv84b          & 0.23  & 4.3  &0.6  \\
 NGC\,4594 &   1.9   &   20.0  &    120      &93.4        &Sad95, NVSS     &-0.21  & 13.0 &0.4  \\
 IC\,3639  &     2   &   35.3  &    20       &87.9        &Sad95, NVSS     &1.23   & 18.0 &1.1  \\
 NGC\,4639 &   1.0   &   16.8  &    0.15     &$<$0.84     &Ho01, FIRST     &       & 16   &0.4  \\
 NGC\,4698 &     2   &   16.8  &    0.23     &$<$1.01     &Ho01, FIRST     &       & 16   &0.4  \\
 NGC\,4725 &     2   &   12.4  &   $<$0.17   &$<$0.97     &Ho01, FIRST     &       & 4.1  &-0.5   \\
 NGC\,4941 &    2    &   16.8  &   4.3       &14.2        &Ulv89           & 0.99  & 3    &-0.4  \\
 NGC\,4939 &    2    &   46.6  &   0.7       &2.3         &Vil90           & 0.99  & 8.3  &1.0  \\
 NGC\,5005 &    2    &   21.3  &   2.7       &24.6        &Vil90           & 1.84  & 10   &0.4  \\
 NGC\,5033 &  1.5    &   18.7  &   5.68      &11.7        &Ho01            & 0.60  & 4.1  &-0.1  \\
 NGC\,5128 &    2    &    4.3  &   6456      &            &Sle94           &       & 3    &-1.5   \\
 NGC\,5135 &    2    &   57.7  &   58.8      &163.2       &Ulv89           & 0.85  & 3    &0.7  \\
 NGC\,5273 &  1.5    &   21.3  &   1.31      &2.13        &Ho01            & 0.40  & 12   &0.4  \\
 NGC\,5395 &    2    &   46.7  &   0.19      &$<$0.92     &Ho01, FIRST     &       & 4.1  &0.7  \\
 NGC\,5427 &    2    &   40.4  &   2.5       &7.54        &Mor99, FIRST    & 0.92  & 3    &0.4  \\
 NGC\,5631 &    2    &   32.7  &   0.29      &  0.43      &Ho01            & 0.33  & 4.1  &0.3  \\
 NGC\,5899 &    2    &   42.8  &   4.0       & 10.35      &Dia09, FIRST    & 0.79  & 3    & 0.4  \\
 NGC\,6221 &    2    &   19.3  &   $<$1.0    &            &Sad95           &       & 11   &0.3  \\
 NGC\,6814 &  1.5    &   25.6  &   20        & 49.7       &Sad95, NVSS     & 0.76  & 5.8  &0.3  \\
 NGC\,6951 &    2    &   24.1  &   9.05      &  25.5      &Ho01            & 0.86  & 3    &-0.1  \\
 MRK\,509  &  1.2    &  143.8  &   1.8       &18.6        &Nef92, NVSS     & 1.94  & 2.7  &1.4  \\
 NGC\,7130 &    2    &   68.7  &   38        &189.7       &Sad95, NVSS     &1.34   & 16   &1.6  \\
 NGC\,7172 &    2    &   37.6  &   11.7      &36.8        &Mor99, NVSS     &0.95   & 6.7  &0.7  \\
 NGC\,7213 &  1.5    &   24.9  &   207       &            &Bra98           &       & 14   &0.6  \\
 NGC\,7314 &  1.9    &   20.8  &   2.7       &31.0        &Mor99, NVSS     &2.03   & 7.4  &0.2  \\
 NGC\,7410 &    2    &   24.8  &   1.4       &36.8        &Con98, NVSS     &2.72   & 14   &0.6  \\
 NGC\,7469 &  1.2    &   67.0  &   21        &$<$0.88     &Sad95, FIRST    &       & 2.9  &0.8  \\
 NGC\,7496 &    2    &   23.1  &   3.8       &            &The00           &       & 15   &0.6  \\
 NGC\,7582 &    2    &   22.0  &   69        &            &Sad95           &       & 10.9 &0.4  \\
 NGC\,7590 &    2    &   22.0  &   $<$0.3    &            &The00           &       & 15   &0.6  \\
 NGC\,7743 &    2    &   24.4  &   3.03      &  5.52      &Ho01            & 0.50  & 3    &0.0  \\

\end{longtable}

 $^{*}$ Radio properties for the nearby nonmaser Seyfert galaxy sample, including 16 Seyfert 1s (type 1-1.5) and 54 Seyfert 2s (type 1.8-2.0).

 Column 1: Source;

 Column 2: Seyfert type, optical classification from Maiolino \& Rieke (1995) or Ho et al. (1997);

 Column 3: Distance in Mpc, taken from Diamond-Stanic et al. (2009);

 Column 4: 6\,cm radio flux density in mJy;

 Column 5: 20\,cm radio flux density in mJy;

 Column 6: References for Cols. 4\&5 (For sources without available 20\,cm data in the literature, their 20\,cm data are taken from the NVSS or FIRST catalog.):
           Ho01: Ho \& Ulvestad (2001); Ulv89: Ulvestad \& Wilson (1989); Mor99: Morganti et al. (1999);
           The00: Thean et al. (2000); Sad95: Sadler et al. (1995); San95: Sandqvist et al. (1995); Sch01: Schmitt et al. (2001);
           Bra98: Bransford  et al. (1998); Ulv84b: Ulvestad \& Wilson (1984b); Vil90: Vila et al. (1990); Sle94: Slee et al. (1984);
           Dia09: Diamond-Stanic et al. (2009); Nef92: Neff \& Hutchings (1992); Con98: Condon et al. (1998);

 Column 7: the spectral index between 6\,cm and 20\,cm, derived from the values in Cols. 4\&5, assuming $S \propto \nu^{-\alpha}$;

 Column 8: RMS values of H$_{2}$O maser data in units of $mJy$, which were taken from MCP and HoME webpages. GBT observations are taken whenever possible and in case of several observations the GBT spectrum with the lowest RMS value was selected;

 Column 9: Estimated upper limits of H$_{2}$O maser luminosity (log\,$L_{\rm H_{2}O}$, in units of $L_{\odot}$) for nonmaser Seyfert galaxies, from
           the RMS value (Column 8), see details in Section 3.2.



\begin{thebibliography}{}

\bibitem[2006]{Baa06} Baan, W. A. \& Kl{\"o}ckner, H. R. 2006, A\&A 449, 559
\bibitem[2005]{Bar05} Barvainis, R., \& Antonucci, R. 2005, ApJ, 628, L89
\bibitem[1991]{Bec91} Becker, R.H., White, R.L. \& Edwards, A.L. 1991, ApJS, 75, 229
\bibitem[1995]{Bec95} Becker, R.H., White, R.L. \& Helfand, D.J. 1995, ApJ, 450, 559
\bibitem[2009]{Ben09} Bennert, N., Barvanis, R., Henkel, C. \& Antonucci, R. 2009, ApJ, 695, 276
\bibitem[2009]{Bor09} Bordalo, V., Plana, H. \& Telles, E. 2009, ApJ, 696, 1668
\bibitem[2009]{Bra09} Braatz, J.A., Condon, J.J., Henkel, C., Lo, K.-Y. \& Reid, M.J. 2009, Astro2010, 23
\bibitem[2008]{Bra08} Braatz, J.A. \& Gugliucci, N.E. 2008, ApJ, 678, 96
\bibitem[1997]{Bra97} Braatz, J.A., Wilson, A.S. \& Henkel, C. 1997, ApJS, 110, 321
\bibitem[1998]{Bra98} Bransford, M.A., Appleton, P.N., Heisler, C.A.,  Norris, R.P. \& Marston, A.P. 1998, ApJ, 497, 133
\bibitem[1977]{Chu77} Churchwell, E., Witzel, A., Huchtmeier, W., Pauliny-Toth, I., Roland, J. \& Sieber, W. 1977, A\&A 54, 969
\bibitem[1983]{Con83} Condon, J.J. 1983, ApJS, 53, 459
\bibitem[2002]{Con02} Condon, J.J., Cotton, W.D. \& Broderick, J.J. 2002, AJ, 124, 675
\bibitem[1998]{Con98} Condon, J.J., Cotton, W.D., Greisen, E.W., Yin, Q.F., Perley, R.A., Taylor, G.B. \& Broderick, J.J. 1998, AJ, 115, 1693
\bibitem[1996]{Con96} Condon, J.J., Helou, G., Sanders, D.B. \& Soifer, B.T. 1996, ApJS, 103, 81
\bibitem[2007]{Coo07} Cooper, N.J., Lister, M.L. \& Kochanczyk, M.D. 2007, ApJS, 171, 376
\bibitem[2008]{Dar08} Darling, J., Brogan, C. \& Johnson, K. 2008, ApJS, 685, 39
\bibitem[2009]{Dia09} Diamond-Stanic, A.M., Rieke, G.H. \& Rigby, J.R. 2009, ApJ, 698, 623
\bibitem[2006]{Fil06} Filho, M.E., Barthel, P.D. \& Ho, L.C. 2006, A\&A 451, 71
\bibitem[2006]{Gal06} Gallimore, J.F., Axon, D.J., O'Dea, C.P., Baum, S.A. \& Pedlar, A. 2006, AJ, 132, 546
\bibitem[1990]{Giu90} Giuricin, G., Mardirossian, F., Mezzetti, M. \& Bertotti, G. 1990, ApJS, 72, 551
\bibitem[2008]{Gre08} Greenhill, L.J., Tilak, A. \& Madejski, G. 2008, ApJ, 686, L13
\bibitem[1991]{Gre91} Gregory, P.C. \& Condon, J.J. 1991, ApJS, 75, 1011
\bibitem[2001]{Gri95} Griffith, M.R. Wright, A.E., Burke, B.F. \& Ekers, R.D. 1995, ApJS, 97, 347
\bibitem[2009]{Guo09} Guo, Q., Zhang, J.S. \& Fan, J.H. 2009, IJMPD,  1809, 1367
\bibitem[1990]{Has90} Haschick, A.D., Baan, W.A., Schneps, M.H. et al., ApJ, 356, 149
\bibitem[1998]{Hen98} Henkel, C., Wang, Y.-P., Falcke, H., Wilson, A.S., \& Braatz, J.A. 1998, A\&A, 335, 463
\bibitem[2005]{Hen05} Henkel, C., Peck, A.B., Tarchi, A., Nagar, N. M., Braatz, J.A., Castangia, P. \& Moscadelli, L. 2005, A\&A, 436, 75
\bibitem[1998]{Her98} Herrnstein, J.R., Greenhill, L.J., Moran, J.M. et al. 1998, ApJ, 497, L69
\bibitem[1997]{Ho97}  Ho, L.C., Filippenko, A.V. \& Sargent, W.L.W. 1997, ApJS, 112, 31
\bibitem[2001]{Ho01}  Ho, L.C. \& Ulvestad, J.S. 2001, ApJS, 133, 77
\bibitem[2009]{Imp08} Impellizzeri, C.M.V., McKean, J.P., Castangia, P., Roy, A.L., Henkel, C., Brunthaler, A. \& Wucknitz, O. 2008, Nature, 456, 927
\bibitem[2010]{Kan10} Kandalyan, R.A. \& M. Al-Zyout 2010, Astrophysics, 53, 475
\bibitem[1999]{Kar99} Kartje, J.F., K{\"o}nigl, A. \& Elitzur M. 1999, ApJ, 513, 180
\bibitem[1999]{Kob99} Kobulnicky, H.A. \& Johnson, K.E. 1999, ApJ, 527, 154
\bibitem[2006]{Kon06} Kondratko, P.T., Greenhill, L.J. \& Moran, J.M. 2006, ApJ, 638, 100
\bibitem[2005]{Lo05}  Lo, K.Y. 2005, ARA\&A, 43, 625
\bibitem[1995]{Mai95} Maiolino, R. \& Rieke, G.H. 1995, ApJ, 454, 95
\bibitem[1995]{Miy95} Miyoshi, M., Moran, J., Herrnstein, J., et al. 1995, Nature, 373, 127
\bibitem[1999]{Mor99} Morganti, R., Tsvetanov, Z.I., Gallimore, J. \& Allen, M.G. 1999, A\&AS, 137, 457
\bibitem[1992]{Nef92} Neff, S.G. \& Hutchings, J.B. 1992, ApJ, 103, 1746
\bibitem[2004]{Neff04} Neff, S.G.,  Ulvestad, J.S. \& Teng, S.H. 2004, ApJ, 611, 186
\bibitem[1994]{Neu94} Neufeld, D. A., Maloney, P. R. \& Conger, S. 1994, ApJ, 436, L127
\bibitem[2007]{Oos07} Oosterloo, T.A., Morganti, R., Sadler, E.M., van der Hulst, T. \& Serra, P. 2007, A\&A 465, 787
\bibitem[2009]{Reid09} Reid, M.J., Braatz, J.A., Condon, J.J., Greenhill, L.J., Henkel, C. \& Lo, K.Y. 2009, ApJ, 695, 287
\bibitem[1995]{Sad95} Sadler, E.M., Slee, O.B., Reynolds, J.E. \& Roy, A.L. 1995, MNRAS, 276, 1373
\bibitem[1995]{San95} Sandqvist, A., Joersaeter, S. \& Lindblad, P.O. 1995, A\&A 295, 58
\bibitem[2001]{Sch01} Schmitt, H.R., Ulvestad, J.S., Antonucci, R.R.J. \& Kinney, A.L. 2001, ApJS, 132, 199
\bibitem[1994]{Sle94} Slee, O.B., Sadler, E.M., Reynolds, J.E. \& Ekers, R.D. 1994, MNRAS, 269, 928
\bibitem[2011]{Tar11} Tarchi, A., Castangia, P., Henkel, C., Surcis, G. \& Menten, K.M. 2011, 525, 91
\bibitem[2002]{Tay02} Taylor, G.B., Peck, A.B., Henkel, C., Falcke, H., Mundell, C.G., O'Dea, C.P., Baum, S.A. \& Gallimore, J.F. 2002, ApJ, 574, 88
\bibitem[2000]{The00} Thean, A., Pedlar, A., Kukula, M.J., Kukula, M.J., Baum, S.A., O'Dea \& Christopher P. 2000, MNRAS, 314, 573
\bibitem[1984]{Ulv84a} Ulvestad, J.S. \& Wilson, A.S. 1984a, ApJ, 278, 544
\bibitem[1984]{Ulv84b} Ulvestad, J.S. \& Wilson, A.S. 1984b, ApJ, 285, 439
\bibitem[1989]{Ulv89} Ulvestad, J.S. \& Wilson, A.S. 1989, ApJ, 343, 659
\bibitem[1990]{Vil90} Vila, M.B., Pedlar, A. \& Davies, R.D. 1990, MNRAS, 242, 379
\bibitem[1992]{Whi92} White, R.L. \& Becker, R.H. 1992, ApJS, 79, 331
\bibitem[1980]{Wil80} Wilson, A.S., Pooley, G.G., Clements, E.D. \& Willis, A.G. 1980, ApJ, 237, 61
\bibitem[1990]{Wri90} Wright, A. \& Otrupcek, R. 1990, PKSCAT90
\bibitem[1994]{Wri94} Wright, A.E., Griffith, M.R., Burke, B.F. \& Ekers, R.D. 1994, ApJS, 91, 111
\bibitem[1996]{Wri96} Wright, A.E., Griffith, M.R., Hunt, A.J., Troup, E., Burke, B.F. \& Ekers, R.D. 1996, ApJs, 103, 145
\bibitem[2010]{Zhang10} Zhang, J.S., Henkel, C., Guo, Q., Wang, H.G. \& Fan, J.H. 2010, ApJ, 708, 1528
\bibitem[2006]{Zhang06} Zhang, J.S., Henkel, C. Kadler, M., Greenhill, L.J., Nagar, N., Wilson, A.S. \& Braatz, J.A. 2006, A\&A, 450, 933
\bibitem[1990]{Zij90} Zijlstra, A., Pottasch, S. \& Bignell, C. 1990, A\&AS, 82, 273

\end{thebibliography}
\end{document}